\documentclass[11pt]{amsart}
\usepackage{hyperref}
\begin{document}

\title[]{A Method for Finding Solutions of the Hermitian Theory
of Relativity which Depend on Three Co-ordinates}
\author{S. Antoci}
\address{Dipartimento di Fisica ``A. Volta'', Via A. Bassi, 6, I-27100
Pavia, Italy}

\thanks{Annalen der Physik, {\bf 44} (1987) 297.}
\subjclass{}%
\keywords{}%

\begin{abstract}
A method is presented, which can generate solutions of the
Hermitian theory of relativity from known solutions of the general
theory of relativity, when the latter depend on three co-ordinates
and are invariant under reversal of the fourth
one.\par\bigskip\noindent
{\bf Eine Methode zum Auffinden von L\"osungen der Hermiteschen
Relativit\"atstheorie, die von drei Koordinaten
abh\"angen}
\par\smallskip\noindent
Inhalts\"ubersicht. Es wird eine Methode vorgestellt, nach der
L\"osungen der Hermiteschen Relativit\"atstheorie aus bekannten
L\"osungen der Allgemeinen Relativit\"atstheorie gewonnen werden
k\"onnen, wenn diese von drei Koordinaten abh\"angen und bei
Umkehrung der vierten invariant bleiben.
\end{abstract}
\maketitle
\section{Introduction}
In recent times several exact solutions for the field equations of
the Hermitian theory or relativity \cite{Einstein} have been found.
Some of them \cite{Antoci1984} have confirmed the result, reached
in 1957 by approximation methods \cite{Treder1957}, that the
theory entails the existence of confined charges, interacting
mutually with forces which do not depend on the distance
\cite{Treder}. Other solutions, in which the antisymmetric part of
the fundamental tensor $g_{ik}$ obeys Maxwell's equations, just
predict the equilibrium positions which are appropriate to
electric charges and currents, so that it is reasonable to believe
that Einstein's Hermitian theory of relativity can provide a
unified description of gravodynamics, chromodynamics and
electrodynamics \cite{Antoci1987}.\par
It was observed that the solutions mentioned above can be
constructed from particular solutions of the field equations of
general relativity by following a certain procedure; the present
paper shows that such an occurrence is not a pure accident, since
a method exists, which can generate solutions of the Hermitian
theory of relativity from given solutions of general relativity,
when the latter depend on three co-ordinates and are invariant
under reversal of the fourth one.\par

\section{A Method for Finding Solutions}
Let us consider a Hermitian fundamental form
$g_{ik}=g_{(ik)}+g_{[ik]}$ and an affine connection
$\Gamma^i_{kl}=\Gamma^i_{(kl)}+\Gamma^i_{[kl]}$ which is
Hermitian with respect to the lower indices.\par
Then the field equations of the Hermitian theory of relativity can
be written as
\begin{eqnarray}\label{1}
g_{ik,l}-g_{nk}\Gamma^n_{il}-g_{in}\Gamma^n_{lk}=0,\\\label{2}
(\sqrt{-g}g^{[is]})_{,s}=0,\\\label{3}
R_{(ik)}(\Gamma)=0,\\\label{4}
R_{[ij],k}(\Gamma)+R_{[ki],j}(\Gamma)+R_{[jk],i}(\Gamma)=0,
\end{eqnarray}
where $g=\det{(g_{ik})}$ and $R_{ik}(\Gamma)$ is the Ricci tensor
\begin{equation}\label{5}
R_{ik}(\Gamma)=\Gamma^a_{ik,a}-\Gamma^a_{ia,k}
-\Gamma^a_{ib}\Gamma^b_{ak}+\Gamma^a_{ik}\Gamma^b_{ab}.
\end{equation}

Consider now a real symmetric tensor $h_{ik}$ corresponding to a
solution of the field equations of general relativity, which
depends on the first three co-ordinates $x^{\lambda}$ and for which
$h_{\lambda 4}=0$; we assume henceforth that Greek indices run
from 1 to 3, while Latin indices run from 1 to 4. Consider also an
antisymmetric purely imaginary tensor $a_{ik}$ which depends on
the first three co-ordinates; assume that its only nonvanishing
components are $a_{\mu 4}=-a_{4 \mu}$. Then form the mixed tensor
\begin{equation}\label{6}
\alpha_i^{~k}=a_{il}h^{lk}=-\alpha^k_{~i},
\end{equation}
where $h^{ik}$ is the inverse of $h_{ik}$, and define the
Hermitian fundamental form $g_{ik}$ as follows:
\begin{eqnarray}\label{7}
g_{\lambda\mu}=h_{\lambda\mu},\\\nonumber
g_{4\mu}=\alpha_4^{~\nu}h_{\nu\mu},\\\nonumber
g_{44}=h_{44}-\alpha_4^{~\mu}\alpha_4^{~\nu}h_{\mu\nu}.\nonumber
\end{eqnarray}
When the three additional conditions
\begin{equation}\label{8}
\alpha^4_{~\mu,\lambda}-\alpha^4_{~\lambda,\mu}=0
\end{equation}
are fulfilled, the affine connection $\Gamma^i_{kl}$ which solves
Eqs. (\ref{1}) has the nonzero components
\begin{eqnarray}\label{9}
\Gamma^{\lambda}_{(\mu\nu)}=\left\{^{~\lambda}_{\mu~\nu}\right\},
\\\nonumber
\Gamma^{\lambda}_{[4\nu]}=\alpha^{~\lambda}_{4~,\nu}
-\left\{^{~4}_{4~\nu}\right\}\alpha^{~\lambda}_4
+\left\{^{~\lambda}_{\rho~\nu}\right\}\alpha^{~\varrho}_4,
\\\nonumber
\Gamma^4_{(4\nu)}=\left\{^{~4}_{4~\nu}\right\},
\\\nonumber
\Gamma^{\lambda}_{44}=\left\{^{~\lambda}_{4~4}\right\}
-\alpha^{~\nu}_4\left(\Gamma^{\lambda}_{[4\nu]}
-\alpha^{~\lambda}_4\Gamma^4_{(4\nu)}\right);
\end{eqnarray}
we indicate with $\left\{^{~i}_{k~l}\right\}$ the Christoffel
connection built with $h_{ik}$; $\Gamma^{\lambda}_{[4\nu]}$
is just written as the covariant derivative of
$\alpha^{~\lambda}_4$ calculated with that connection. We form now
the Ricci tensor $R_{ik}(\Gamma)$. When Eqs. (\ref{2}), i.e., in
our case, the single equation
\begin{equation}\label{10}
(\sqrt{-h}~\alpha^{~\lambda}_4 h^{44})_{,\lambda}=0,
\end{equation}
and the additional conditions, expressed by Eqs. (\ref{8}), are
satisfied, the components of $R_{ik}(\Gamma)$ can be written as
\begin{eqnarray}\label{11}
R_{\lambda\mu}=S_{\lambda\mu},
\\\nonumber
R_{4\mu}=\alpha^{~\nu}_4S_{\nu\mu}+\left(\alpha^{~\nu}_4
\left\{^{~4}_{4~\nu}\right\}\right)_{,\mu},
\\\nonumber
R_{44}=S_{44}-\alpha^{~\mu}_4\alpha^{~\nu}_4S_{\mu\nu},
\end{eqnarray}
where $S_{ik}$ is the Ricci tensor built with
$\left\{^{~i}_{k~l}\right\}$. $S_{ik}$ is zero when $h_{ik}$ is a
solution of the field equations of general relativity, as
supposed; therefore, when Eqs. (\ref{8}) and (\ref{10}) hold, the
Ricci tensor, defined by Eqs. (\ref{11}), satisfies Eqs. (\ref{3})
and (\ref{4}) of the Hermitian theory of relativity.
\section{Conclusion}
We can resume our result as follows: let $h_{ik}$ be the metric
tensor for a solution of the field equations of general relativity
which does not depend on, say, the fourth co-ordinate, and for
which $h_{\lambda 4}=0$. Let $a_{ik}$ be an antisymmetric, pure
imaginary tensor, which depends on the first three co-ordinates;
$a_{4\mu}=-a_{\mu 4}$ are its only nonvanishing components. When
$a_{ik}$ obeys the field equation (\ref{10}) and the additional
conditions (\ref{8}), the fundamental form $g_{ik}$, defined by
Eqs. (\ref{7}), provides a solution to the field equations of the
Hermitian theory of relativity.\par
    The task of solving Eqs. (\ref{1})-(\ref{4}) reduces, under
the circumstances considered here, to the simpler task of solving
Eqs. (\ref{8}) and (\ref{10}) for a given $h_{ik}$; the particular
solutions mentioned in the Introduction are physically meaningful
examples of solutions built in this way.\par
    We notice finally that the method proposed here applies also
to Schr\"odin\-ger's purely affine theory \cite{Schroedinger}.
\bigskip

Aknowledgements. I am indebted to Prof. E. Kreisel and to Prof.
H.-J. Treder for helpful discussions and for continuous support.

\bibliographystyle{amsplain}

\end{document}